\documentclass[conference]{IEEEtran}

\usepackage{amsmath,amsfonts}
\usepackage{amsthm}

\usepackage{algorithmic}
\usepackage{array}
\usepackage{textcomp}
\usepackage{stfloats}
\usepackage{url}
\usepackage{verbatim}
\usepackage{graphicx}
\usepackage{amssymb}
\usepackage{color,soul}
\usepackage{amsthm} 
\usepackage{bbm}
\usepackage{algorithm}
\usepackage{caption}
\usepackage{bm}
\usepackage{subcaption}
\usepackage{cite}
\usepackage[top=1.8cm,bottom=4.2cm]{geometry}
\def\BibTeX{{\rm B\kern-.05em{\sc i\kern-.025em b}\kern-.08em
    T\kern-.1667em\lower.7ex\hbox{E}\kern-.125emX}}

\usepackage{hyperref}

\begin{document}
\title{
Delay-Aware Large-Small Model Collaboration over LEO Satellite Networks
}

\author{
Mingyu~Guo$^{* \dagger}$,~Wen~Wu$^{*}$,~Ying~Wang$^{\dagger}$,~Songge~Zhang$^{*}$,~and~Liang~Li$^{*}$
\\
{\normalsize \textsuperscript{*}Frontier Research Center, Pengcheng Laboratory, Shenzhen, China}\\
{\normalsize \textsuperscript{$\dagger$}School of Information and Communication Engineering, Beijing University of Posts and Telecommunications}\\
{\normalsize Email:\{guomy01, wuw02, zhangsg, lil03\}@pcl.ac.cn$^{*}$, wangying@bupt.edu.cn$^{\dagger}$}
}

\maketitle

\begin{abstract}
In this paper, we introduce a delay-aware large-small model collaboration scheme for low Earth orbit (LEO) satellite networks, which can balance the computational load among satellites and the communication load across inter-satellite links.
Specifically, computational resource constrained remote sensing satellites are responsible for data collection and local processing using small models, 
while collaborating with computing satellites that provide large model processing.
To minimize the service delay, we formulate a joint optimization problem for offloading decision and routing strategy design, which is transformed into a decentralized partially observable Markov decision process.
To solve the problem, we develop a multi-agent reinforcement learning (MARL)-based algorithm with offline policy training and online bisection search.
The offline trained policy determines routing strategies, while online bisection search iteratively adjusts the offloading decisions.
Simulation results demonstrate that the proposed scheme can reduce the service delay by up to 31.85\% compared with the benchmarks.
\end{abstract}

\section{Introduction}
\IEEEPARstart{L}ow Earth orbit (LEO) satellites are emerging as a key component of sixth-generation (6G) networks due to their ability to support communication, sensing, and computing services~\cite{sxm-2022}.
Recent advances in satellite hardware, such as onboard GPUs and artificial intelligence (AI) accelerators, have made on-orbit intelligent computing increasingly feasible for LEO satellite networks~\cite{ZSG-WCSP}.
{
For example, Google has recently explored space based AI infrastructure through Project Suncatcher, which envisions solar powered LEO satellite constellations equipped with TPUs and interconnected by free space optical links. 
}
On-orbit intelligent computing is particularly important for remote sensing services, where a single remote sensing satellite can generate terabyte-scale data on a daily basis~\cite{sharma-2021}.  
Instead of transmitting raw sensing data to ground stations for offline processing, on-orbit computing enables satellites to process data directly in space and transmit only computation results, thereby reducing satellite–ground communication overhead and service delay~\cite{LK-2026}. 
As a result, LEO satellite networks are evolving from simple data relays into distributed platforms for space-based intelligent computing~\cite{QY-2025}.

With the rapid development of large models~\cite{wu4}, the number of model parameters has increased substantially~\cite{wu10}, which may lead to prolonged service delays on satellites with limited computing resources.
Existing studies have investigated model compression techniques, such as model pruning~\cite{modelpruning} and weight sparsification~\cite{weightsparse}, to reduce service delay.
These methods mainly reduce the number of model parameters and lower computational complexity.
Wang \emph{et al.} proposed a sparse quantized CNN with a co-designed hardware accelerator for efficient real-time onboard remote sensing~\cite{WT-2025}.
Wei \emph{et al.} developed a differentiable neural architecture search based channel pruning method to compress CNNs for onboard remote sensing scene classification~\cite{WX-2022}.
However, these compression-based techniques may inevitably degrade model accuracy.

Large–small model collaboration is a promising technique to mitigate the accuracy degradation caused by model compression~\cite{wu10}.
This technique can effectively utilize onboard computational resources while reducing service delay.
However, existing studies typically assume homogeneous computational capabilities across satellites, thereby overlooking the inherent heterogeneity of satellite resources.
Moreover, multi-satellite collaborative processing faces significant challenges under bursty task arrivals, which can lead to congestion over inter-satellite links (ISLs).
Therefore, it is essential to develop a collaborative low-latency on-orbit processing scheme for LEO satellite networks.

In this paper, we propose a large-small model collaboration scheme for LEO satellite networks.
Specifically, computation is collaboratively executed by remote sensing satellites and computing satellites, thereby addressing the limited computing capability of remote sensing satellites.
The satellites collaborate to execute the task, with feature extraction and residual mapping transmitted via ISLs for synchronous processing.
We reduce the transmission delay and prevent congestion on ISLs through an intelligent routing strategy. 
We formulate an optimization problem to minimize service delay by identifying the optimal allocation ratio and routing subject to accuracy constraints.
To solve this problem, we model it as a decentralized partially observable Markov decision process, in which satellites are represented as agents that make routing decisions based on the LEO network state.
We then design a bisection search (BS)-based multi-agent reinforcement learning (MARL) algorithm to find the optimal allocation ratio and routing strategy.
Simulation results demonstrate that the proposed scheme can reduce service delay by up to 30\% compared to the other benchmarks.
The main contributions of this paper can be summarized as follows.
\begin{itemize}
    \item  {We propose a large-small model collaboration scheme for LEO satellite networks;}
    \item We formulate a service delay minimization problem to jointly optimize the offloading decisions and routing strategies;
    \item {We propose a BS-MARL algorithm to determine the optimal decision variables.}
\end{itemize}

The remainder of this paper is organized as follows. 
Section~\ref{sec:proposed_scheme} presents the proposed scheme and delay analysis.
Section~\ref{sec:problem_formulation} presents problem formulation.
Section~\ref{sec:opt_alg} details the  proposed algorithm.
Section~\ref{sec:sim_res} presents the
simulation results.
Finally, Section~\ref{sec:conclusion} concludes the paper.

\begin{figure}[t]
  \centering
  \includegraphics[width=0.45\textwidth]{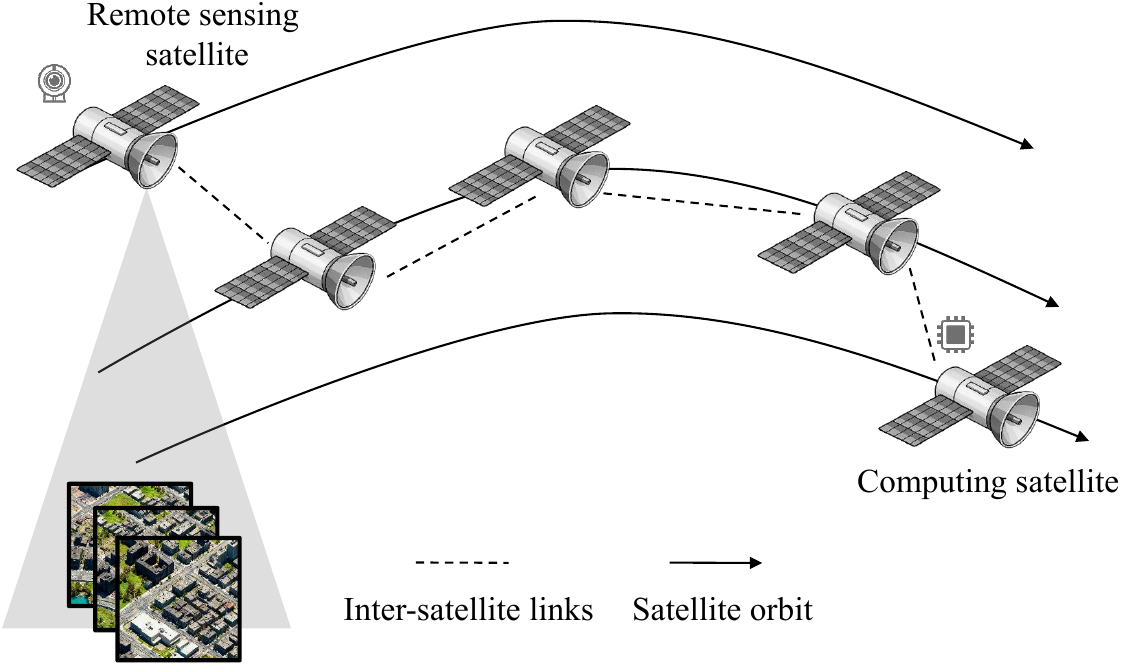}
  \caption{ Considered scenario.}
  \label{system_model}
\end{figure}

\section{Proposed Scheme}
\label{sec:proposed_scheme}

\subsection{Considered Scenario}
In this work, we consider a LEO satellite network consisting of heterogeneous satellites, in which large-small model collaboration is performed, as shown in Fig.~\ref{system_model}.
Let $\mathcal{S}_{\mathrm{RS}}$, $\mathcal{S}_{\mathrm{COM}}$, and $\mathcal{S}_{\mathrm{CPT}}$ denote the sets of remote sensing satellites, relay satellites, and computing satellites, respectively, with $|\mathcal{S}_{\mathrm{RS}}|=N_{\mathrm{RS}}$, $|\mathcal{S}_{\mathrm{COM}}|=N_{\mathrm{COM}}$, and $|\mathcal{S}_{\mathrm{CPT}}|=N_{\mathrm{CPT}}$.
Satellites in $\mathcal{S}_{\mathrm{RS}}$ undertake sensing tasks using a small model and provide local computing capability.
Satellites in $\mathcal{S}_{\mathrm{COM}}$ serve as communication relay nodes to support ISLs.
Satellites in $\mathcal{S}_{\mathrm{CPT}}$ process sensing tasks using a large model and provide computational resources.
Through the collaboration between remote sensing and computing satellites, vision-based inference is carried out for downstream applications, and the final results are transmitted to the ground station via downlink.

The considered scenario enables on-orbit real-time processing of remote sensing data, with only the results transmitted to the ground station, thus reducing the need to send raw data.
To reduce service delay and achieve load balancing, a large-small model collaboration scheme is proposed in the following subsection.

\subsection{Large-Small Model Collaboration Scheme}
We propose a large-small model collaboration scheme for LEO satellite networks.
Along the ISLs, transmitted payloads are categorized as data packets and model packets. 
Data packets transmit extracted feature data and residual mapping data from the remote sensing satellites to the computing satellite, while model packets deliver model updating information from the computing satellite back to the remote sensing satellites.

\subsubsection{Collaborative Inference}For each remote sensing satellite $i \in \mathcal{S}_{\mathrm{RS}}$, let $f_i$, $p_i$, and $b_i$ denote the number of image frames, the number of pixels per frame, and the number of quantization bits per pixel, respectively. The total onboard image data volume is $D_i^{\mathrm{img}} = b_i f_i p_i$.
A fraction $\alpha_i$ of the frames is offloaded via ISLs to a computing satellite for large-model processing. Let $\Psi_i$ denote the set of offloaded frames, where $\left|\Psi_i\right| = \alpha_i f_i$.
The remaining fraction, $(1-\alpha_i)f_i$, is processed locally using the small model on satellite $i$.

To reduce the communication load, the remote sensing satellite extracts features from the frames in $\Psi_i$~\cite{ZSH-2025} and transmits them to the computing satellite.
Let $F_{\mathrm{p}}$ denote the average feature data volume per frame~\cite{YWH-2024}. Then the total transmitted feature volume is $D_i^{\mathrm{f}} = \alpha_i f_i F_{\mathrm{p}}$.

To mitigate the accuracy loss caused by detail reduction, residual mapping data are further transmitted for a subset of offloaded frames. Let $\Phi_i \subseteq \Psi_i$ denote the set of frames with residual transmission, where $\left|\Phi_i\right| = \beta_i f_i$.
Let $q_{i}^{m}$ denote the quantization parameter for the residual data of frame $m \in \Phi_i$, which is selected  from a set of discrete values $\Omega$.
Then the total residual data volume is $D_i^{\mathrm{res}} = \sum_{m \in \Phi_i} p_i q_{i}^{m}$.
Accordingly, the total data packet size transmitted from remote sensing satellite $i$ to a computing satellite is $D_i^{\mathrm{d}} = D_i^{\mathrm{f}} + D_i^{\mathrm{res}}$.
    
\subsubsection{Model Updating}
\begin{figure}[t]
  \centering
  \includegraphics[width=0.45\textwidth]{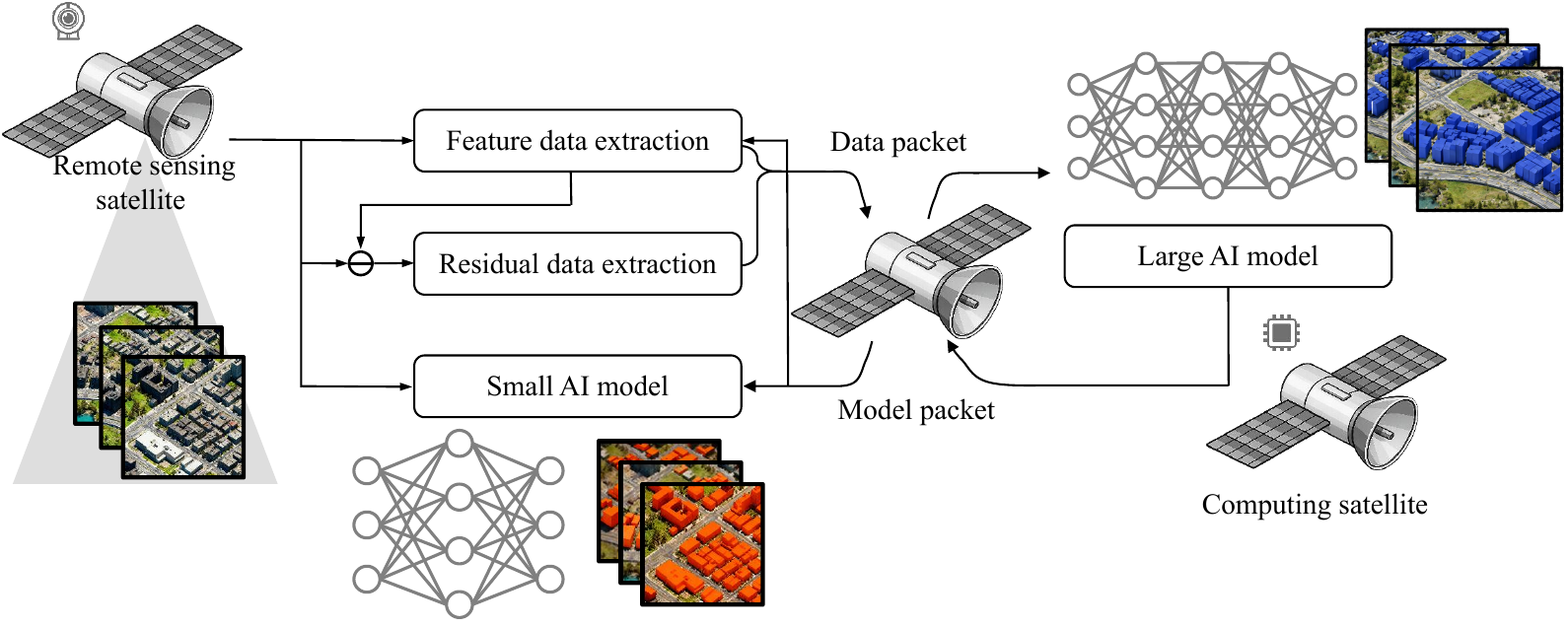}
  \caption{Paradigm for the large-small model collaboration scheme for LEO satellite networks.}
  \label{framework}
\end{figure}

Constrained by limited computational capability and power resources, the small model deployed on the remote sensing satellite is incapable of performing model updates. 
In contrast, a large amount of remote sensing image features and residual mapping data are transmitted to the computing satellite via the ISLs. 
As shown in Fig.~\ref{framework}, leveraging its ample computational resources, the computing satellite can generate and transmit the model packet for updating, thereby improving the vision-based classification accuracy of the small model deployed on the remote sensing satellites.
Let $D_i^{\mathrm{m}}$ denote the model packet size for updating the small model on remote sensing satellite $i$.
Then $D_{\min}^{\mathrm{m}} \leq D_i^{\mathrm{m}} \leq D_{\max}^{\mathrm{m}}$, where $D_{\min}^{\mathrm{m}}$ and $D_{\max}^{\mathrm{m}}$ denote the minimum and maximum model update overhead, respectively.

\subsubsection{Intelligent Routing}
The limited number of onboard communication terminals constrains the communication capacity of each satellite, such that each satellite can establish ISLs with only a limited subset of its neighboring satellites. Therefore, in the considered satellite communication network, each satellite is treated as a node in the routing process. Let $\mathcal{S}=\mathcal{S}_{\mathrm{RS}}\cup\mathcal{S}_{\mathrm{COM}}\cup\mathcal{S}_{\mathrm{CPT}}$ denote the set of all satellites, and let $\mathcal{E}$ denote the set of ISLs. For any link $e=(u,v)\in\mathcal{E}$, its two endpoints satisfy $u,v\in\mathcal{S}$.
The data packet, which contains feature information and residual mapping data, is transmitted from a remote sensing satellite to a computing satellite through ISLs. For each remote sensing satellite $i\in\mathcal{S}_{\mathrm{RS}}$, the routing path of its data packet is denoted by $\mathbf{r}_i^{\mathrm{d}}=\left(r_{i,0}^{\mathrm{d}}, r_{i,1}^{\mathrm{d}}, \dots, r_{i,H_i^{\mathrm{d}}}^{\mathrm{d}}\right)$, where $H_i^{\mathrm{d}}$ denotes the hop count of the data packet route, $r_{i,0}^{\mathrm{d}}=i$, $r_{i,H_i^{\mathrm{d}}}^{\mathrm{d}}\in\mathcal{S}_{\mathrm{CPT}}$, and $r_{i,h}^{\mathrm{d}}\in\mathcal{S}_{\mathrm{COM}}$ for $h\in \left[1,H^{\text{d}}_{i}-1\right]$.
Accordingly, the set of all data packet routing paths is written as $\mathcal{R}^{\mathrm{d}}=\left\{\mathbf{r}_i^{\mathrm{d}}\mid i\in\mathcal{S}_{\mathrm{RS}}\right\}$.
The routing path of the model packet destined for remote sensing satellite $i$ is denoted by $\mathbf{r}_i^{\mathrm{m}}=\left(r_{i,0}^{\mathrm{m}}, r_{i,1}^{\mathrm{m}}, \dots, r_{i,H_i^{\mathrm{m}}}^{\mathrm{m}}\right)$, where $H_i^{\mathrm{m}}$ denotes the hop count of the model packet route, $r_{i,0}^{\mathrm{m}}\in\mathcal{S}_{\mathrm{CPT}}$, $r_{i,H_i^{\mathrm{m}}}^{\mathrm{m}}=i$, and $r_{i,h}^{\mathrm{m}}\in\mathcal{S}_{\mathrm{COM}}$ for $h\in \left[1,H^{\text{m}}_{i}-1\right]$. 
The set of all model packet routing paths is written as $\mathcal{R}^{\mathrm{m}}=\left\{\mathbf{r}_i^{\mathrm{m}}\mid i\in\mathcal{S}_{\mathrm{RS}}\right\}$.

\section{System Model}

\subsection{Communication Delay}
For any link $e=(u,v)\in\mathcal{E}$, let $C_e^{\mathrm{ISL}}$ denote the achievable transmission rate of link $e$~\cite{ZD-2021}. Let $n_e^{\mathrm{m}}$ and $n_e^{\mathrm{d}}$ denote the numbers of model packets and data packets transmitted over link $e$, respectively. 
For the packets associated with remote sensing satellite $i\in\mathcal{S}_{\mathrm{RS}}$, the transmission times of the model packet and data packet over link $e$ are given by $t_{i,e}^{\mathrm{tran,m}}=D_i^{\mathrm{m}} n_e^{\mathrm{m}}/C_e^{\mathrm{ISL}}$ and $t_{i,e}^{\mathrm{tran,d}}=D_i^{\mathrm{d}} n_e^{\mathrm{d}}/C_e^{\mathrm{ISL}}$, respectively~\cite{QXH-2023}. 

Let $\mathcal{E}_i^{\mathrm{m}}$ and $\mathcal{E}_i^{\mathrm{d}}$ denote the sets of links along the routing paths of the model packet and data packet associated with remote sensing satellite $i$, respectively. 
Then, the total transmission time of the model packet $T_i^{\mathrm{tran,m}}$ and data packet $T_i^{\mathrm{tran,d}}$ are given by $\sum_{e\in\mathcal{E}_i^{\mathrm{m}}} t_{i,e}^{\mathrm{tran,m}}$ and $\sum_{e\in\mathcal{E}_i^{\mathrm{d}}} t_{i,e}^{\mathrm{tran,d}}$.
The propagation delay over link $e$ is $t_e^{\mathrm{prop}}=d_e/c$, where $d_e$ is the length of link $e$ and $c$ is the speed of light. Accordingly, the total propagation times of the model packet $T_i^{\mathrm{prop,m}}$ and data packet $T_i^{\mathrm{prop,d}}$ are given by $\sum_{e\in\mathcal{E}_i^{\mathrm{m}}} t_e^{\mathrm{prop}}$ and $\sum_{e\in\mathcal{E}_i^{\mathrm{d}}} t_e^{\mathrm{prop}}$.

\subsection{Computation Delay}

In the proposed large-small model collaboration scheme for LEO satellite networks, the computation delay consists of inference delay and relay processing delay. 
A fraction $\alpha_i$ of frames is selected for feature extraction.
Let $V_i$~$\text{(cycle/bit)}$ denote the required computational workload for feature extraction per bit, $F_i$~$\text{(cycle/s)}$ denote the computing capacity of remote sensing satellite $i$, and $\gamma_i$ its hardware utilization factor. 
Then the feature extraction time is $T_i^{\mathrm{f}}=\alpha_i D_i^{\mathrm{img}} V_i / (\gamma_i F_i)$.

The remaining fraction, $1-\alpha_i$, is processed locally on the small model of satellite $i$. Let $P_i^{\mathrm{loc}}$ denote the number of operations required for one forward pass of the small model on each frame. Then the local inference time is $T_i^{\mathrm{loc}}=(1-\alpha_i) f_i P_i^{\mathrm{loc}} / (\gamma_i F_i)$.
Let $k\in\mathcal{S}_{\mathrm{CPT}}$ denote the computing satellite associated with remote sensing satellite $i$. After the data packets are offloaded, the large model inference time on computing satellite $k$ is $T_i^{\mathrm{lar}}=\alpha_i f_i P_i^{\mathrm{lar}} / (\gamma_{k} F_{k})$, where $P_i^{\mathrm{lar}}$ denotes the number of operations required for one forward pass of the large model on each offloaded frame.

Satellites along the routing path also incur relay processing delay when forwarding packets.
For each relay satellite $j\in\mathcal{S}_{\mathrm{COM}}$, let $m_j^{\mathrm{d}}$ and $m_j^{\mathrm{m}}$ denote the numbers of data packets and model packets relayed by satellite $j$, respectively.
Let $\xi_j$ denote the processing time per bit at relay satellite $j$.
Then, for the packets associated with remote sensing satellite $i$, the relay processing times at satellite $j$ are given by $t_{i,j}^{\mathrm{proc,d}}=\xi_j D_i^{\mathrm{d}} m_j^{\mathrm{d}}$ and $t_{i,j}^{\mathrm{proc,m}}=\xi_j D_i^{\mathrm{m}} m_j^{\mathrm{m}}$~\cite{QXH-2023}.
Accordingly, the total relay processing times of the data packet and model packet are given by $T_i^{\mathrm{proc,d}}=\sum_{h=1}^{H_i^{\mathrm{d}}-1} t_{i,r_{i,h}^{\mathrm{d}}}^{\mathrm{proc,d}}$ and $T_i^{\mathrm{proc,m}}=\sum_{h=1}^{H_i^{\mathrm{m}}-1} t_{i,r_{i,h}^{\mathrm{m}}}^{\mathrm{proc,m}}$, respectively.

\subsection{Delay Analysis}
The large-small model collaboration scheme consists of three phases.
First, the model packet is transmitted to the remote sensing satellite.
Second, the remote sensing satellite performs feature extraction of the images.
Third, the data packets are subsequently transmitted to the large model on the computing satellite. 
The remaining remote sensing images are processed locally by the small model on the remote sensing satellite.

The total transmission delay of the model packet transmitted from the computing satellite and received by the remote sensing satellite over the ISL is given by
\begin{equation}\label{tmodel}
    \begin{aligned}
        T^{\text{m}}_{i} 
        = 
T_{i}^{\text{proc,m}}
+
T_{i}^{\text{tran,m}}
+
T_{i}^{\text{prop,m}}.
    \end{aligned}
\end{equation}
The total transmission delay of the data packet sent from the remote sensing satellite and received by the computing satellite over the ISL can be expressed as
\begin{equation}\label{tdata}
    \begin{aligned}
        T^{\text{d}}_{i} 
        = T_{i}^{\text{proc,d}}
         + T_{i}^{\text{tran,d}}
         +
         T_{i}^{\text{prop,d}}.
    \end{aligned}
\end{equation}
The service delay for processing the remote sensing data task on the remote sensing satellite $i$ can be expressed as
\begin{equation}
    \label{T_total}
    \begin{aligned}
        T_{i}^{\text{total}}
    =
    T^{\text{m}}_{i}
    +
    T^{\text{f}}_{i} 
    +
    &\max\left[ T^{\text{d}}_{i}+T^{\text{lar}}_{i}, T^{\text{loc}}_{i}\right],
    \end{aligned}
\end{equation}
where $T_i^{\mathrm{d}} + T_i^{\mathrm{lar}}$ denotes the processing delay of large model processing, while $T_i^{\mathrm{loc}}$ denotes the delay of local processing. Since these two procedures are executed in parallel after feature extraction, the overall completion time is determined by their maximum.

\section{Problem Formulation}
\label{sec:problem_formulation}

\subsection{Problem Formulation}
An optimization problem is formulated to minimize the service delay under the proposed large-small model collaboration scheme. 
The objective is to determine the offloading decisions and routing strategies, subject to mean average precision\footnote{$m{\rm AP}$ is defined as the area under the precision–recall curve, which evaluates the overall classification performance~\cite{ZSH-2025}.} ($m{\rm AP}$).
For the large model on computing satellite, the $m{\rm AP}$ depends on the quality of image features and residual data from the remote sensing satellite, 
$m{\rm AP}_{i}^{L} = g(\beta_{i}, q_i^{m}), \forall m \in \Phi_i$~\cite{YWH-2024}.
For small models on remote sensing satellite $i$, the $m{\rm AP}$ is determined by the model packets from the computing satellite, 
$m{\rm AP}_{i}^{S} = h(D_{i}^{\text{m}})$.
The expressions of functions $g(\cdot)$ and $h(\cdot)$ can be fitted from experimental data~\cite{ZSH-2025}.

According to~\cite{ZSH-2025}, when $\alpha_i=\beta_i$ and 
$q_i^1 = q_i^2 = \cdots = q_i^m=\bar{q}_i,\ \forall m \in \Phi_i$, the $m{\rm AP}$ of the large model is maximized.
The optimization problem is formulated as follows
\begin{subequations}
\label{problem_formulation}
    \begin{align}
    &\min_{\alpha_{i},
    D_{i}^{\text{m}},
    \mathcal{R}^{\text{d}}, \mathcal{R}^{\text{m}}} \quad \frac{\sum_{i=1}^{N_{\text{RS}}} T_{i}^{\text{total}}}{N_{\text{RS}}}
    \label{PF_a}
    \\
    \text{s.t.} \quad  &m{\rm AP}_{i}^{L}= g(\beta_{i}, \bar{q}_i) \geq m{\rm AP}_{\rm min}, 
    \label{PF_b}
    \\
    &m{\rm AP}_{i}^{S} = h(D_{i}^{\text{m}}) \geq m{\rm AP}_{\rm min},
    \label{PF_d}
    \\
    & D_{\text{min}}^{\text{m}} \leq D_{i}^{\text{m}} \leq D_{\text{max}}^{\text{m}},
    \label{PF_g}
    \\ 
    & r_{i,m}^{\mathrm{d}}\neq r_{i,n}^{\mathrm{d}},
\forall r_{i,m}^{\mathrm{d}},r_{i,n}^{\mathrm{d}}\in \mathbf{r}_i^{\mathrm{d}},
\label{PF_j}
\\
& r_{i,m}^{\mathrm{m}}\neq r_{i,n}^{\mathrm{m}},
\forall r_{i,m}^{\mathrm{m}},r_{i,n}^{\mathrm{m}}\in \mathbf{r}_i^{\mathrm{m}}.
\label{PF_k}
\end{align}
\end{subequations}

Constraint \eqref{PF_d}, \eqref{PF_b} specifies the minimum $m{\rm AP}$ for each task, \eqref{PF_g} constrains the overhead of updating the small model, and \eqref{PF_j}, \eqref{PF_k} prohibit repeated selection of the same satellite during data and model packet routing.

The above optimization problem is non-trivial to solve due to the following three challenges.
First, the problem \eqref{problem_formulation} is a mixed-integer nonlinear program and is NP-hard.
Second, as the $m{\rm AP}$ functions $g(\cdot)$ and $h(\cdot)$ are obtained via experimental fitting, the problem's convexity cannot be guaranteed~\cite{ZSH-2025}. 
Third, the optimization variables for each task are highly coupled and each task's routing affects the ISLs in LEO satellite networks.

\subsection{Problem Decomposition}
The strong coupling of the optimization variables leads to significant challenges in directly solving problem \eqref{problem_formulation}.
To address this issue, we adopt an alternating optimization approach. Specifically, when the task allocation ratio $\alpha_i$ is fixed, the offloading related variables can be separated from the routing decision, and problem~\eqref{problem_formulation} can be decomposed into the offloading subproblem and the routing subproblem.
The allocation ratio and the routing strategy are then optimized alternately until convergence, yielding an efficient solution to the original problem.
In the offloading subproblem, the objective is to minimize the transmission volume of data and model packets, which jointly determine the $m{\rm AP}$ of the large model on the computing satellite and the small model on the remote sensing satellite. 
The data packet design subproblem is formulated as
\begin{subequations}
\label{sub_problem_1d}
    \begin{align}
    &\min_{\bar{q}_i} \quad D_{i}^{\text{d}},
    \label{SP1_a}
    \\
    \text{s.t.} \quad 
    &\bar{q}_i \in \Omega, \quad \forall i \in \mathcal{S}_{\rm RS}.
    \\
    &\eqref{PF_b}.    
\end{align}
\end{subequations}
{The model packet design subproblem is given by}
\begin{subequations}
\label{sub_problem_1m}
    \begin{align}
    &\min \quad D_{i}^{\text{m}},
    \label{SP1_e}
    \\
    \text{s.t.}\quad&  
\eqref{PF_d},~\text{and}~\eqref{PF_g}.
\end{align}
\end{subequations}

Subproblem \eqref{sub_problem_2} focuses on designing the routing strategies for data and model packets over the ISL, denoted by $\mathcal{R}^{\text{d}}$ and $\mathcal{R}^{\text{m}}$. 
Its objective is to minimize the service delay of the remote sensing data processing task, as formulated below

\begin{subequations}
\label{sub_problem_2}
    \begin{align}
    &\min_{ 
    \mathcal{R}^{\text{d}}, \mathcal{R}^{\text{m}}} \quad \frac{\sum_{i=1}^{N_{\text{RS}}} T_{i}^{\text{total}}}{N_{\text{RS}}},
    \label{SP2_a}
    \\ 
    \text{s.t.} \quad &
\eqref{PF_j},~\text{and}~\eqref{PF_k}.
\end{align}
\end{subequations}




\subsection{MDP-Based Problem Transformation}
{
Based on the above problem decomposition, the offloading subproblems can be solved to determine the model packet size $D_i^{\mathrm m}$ and the data packet size $D_i^{\mathrm d}$ for each remote sensing satellite $i\in\mathcal{S}_{\mathrm{RS}}$.
Then, subproblem~\eqref{sub_problem_2}, which determines the routing strategies $\mathcal{R}^{\mathrm d}$ and $\mathcal{R}^{\mathrm m}$ over the time-varying ISL network, can be viewed as a multi-agent sequential decision-making problem.
Since the routing decisions of different satellites are coupled through shared ISLs and relay reuse, we formulate the routing process as a decentralized partially observable Markov decision process (Dec-POMDP), denoted by
$
\bigl(\mathcal{N},\mathcal{X},\{\mathcal{A}_n\}_{n\in\mathcal{N}},P,\{\mathcal{O}_n\}_{n\in\mathcal{N}},\varrho\bigr)
$.
The definition is explained as follows
\begin{enumerate}
    \item[$\bullet$] $\mathcal{N}$ denotes the agent set. 
    Let $\mathcal{N}=\mathcal{S}$, where each satellite is regarded as an agent.

    \item[$\bullet$] $\mathcal{X}$ denotes the global state space. 
    The global state $\mathbf{x}\in\mathcal{X}$ characterizes the state space of the multi-agent system, integrating the individual states of all intelligent agents together with the environmental states.

    \item[$\bullet$] $\{\mathcal{A}_n\}_{n\in\mathcal{N}}$ denotes the collection of local action spaces. 
    The local action space of agent $n\in\mathcal{N}$ is defined as
    $\mathcal{A}_n=\{a_{\mathrm{u}},a_{\mathrm{d}},a_{\mathrm{l}},a_{\mathrm{r}}\}$,
    where $a_{\mathrm{u}}$ and $a_{\mathrm{d}}$ denote forwarding the packet to the preceding and succeeding satellites in the same orbit, respectively, and $a_{\mathrm{l}}$ and $a_{\mathrm{r}}$ denote forwarding the packet to satellites in the adjacent left and right orbits, respectively. 
    Invalid actions are masked according to the current ISL establishment status. 
    Accordingly, the joint action space is $\mathcal{A}_{\mathrm{total}}=\prod_{n\in\mathcal{N}}\mathcal{A}_n$. 

    \item[$\bullet$] $P$ denotes the state transition probability. 
    Let $\mathbf{a}\in\mathcal{A}_{\mathrm{total}}$ denote the joint action of all agents. 
    The transition probability is $P(\mathbf{x}' | \mathbf{x},\mathbf{a})$, which characterizes the evolution from the current global state $\mathbf{x}$ to the next global state $\mathbf{x}'$ after executing the joint action $\mathbf{a}$. 

   \item[$\bullet$] $\{\mathcal{O}_n\}_{n\in\mathcal{N}}$ denotes the collection of local observation spaces. 
    The local observation of agent $n$ is denoted by $o_n\in\mathcal{O}_n$. 
    Specifically, $o_n$ includes the locally available routing information, such as whether the satellite currently carries a packet, the packet type, task information, neighboring ISL establishment status, distance to the destination satellite, reuse status of adjacent relays and ISLs, accumulated delay, and whether the packet has reached its destination. 

    \item[$\bullet$] $\varrho$ denotes the global reward function. 
    The global reward function $\varrho(\mathbf{x},\mathbf{a})$ is 
    $
        r_0(\mathbf{x},\mathbf{a})
        +
        \delta r_1(\mathbf{x},\mathbf{a})$,
    where $r_0(\mathbf{x},\mathbf{a})$ is the reward for successful task completion and $\delta>0$ is a weighting factor. 
    The delay aware reward term $r_1(\mathbf{x},\mathbf{a})$ is $
        \sum_{i=1}^{N_{\mathrm{RS}}}
        \frac{1}{1+\exp\left((\bar{T}_{i}^{\mathrm{total}}-1)/\mu\right)}$, where $\bar{T}_{i}^{\mathrm{total}}$ denotes the normalized service delay of the task generated by remote sensing satellite $i$, and $\mu>0$ is a shaping parameter controlling the sensitivity of the reward to delay. 
\end{enumerate}
}

\begin{algorithm}[t!]\label{alg1}
{\small
\caption{BS-MARL Algorithm}
{\bf Input:} {System and network parameters, trained MARL routing policy $\pi_{\phi}$, maximum bisection iterations $K$};

{\bf Output:} {Allocation ratio $\alpha_i$, routing paths $\mathcal{R}^{\mathrm d}$ and $\mathcal{R}^{\mathrm m}$, and service delay $T_i^{\mathrm{total}}$};

\begin{algorithmic}[1]
\STATE Initialize $\alpha_i^{\mathrm{low}}$ and $\alpha_i^{\mathrm{up}}$;
\FOR{$k=1,\cdots,K$}
    \STATE Set $\alpha_i^{k}=(\alpha_i^{\mathrm{low}}+\alpha_i^{\mathrm{up}})/2$;
    \STATE Obtain $D_i^{\mathrm d}$ and $D_i^{\mathrm m}$ via~\eqref{sub_problem_1d} and~\eqref{sub_problem_1m};
    \STATE Invoke the trained MARL policy $\pi_{\phi}$ to generate data packet routing paths $\mathcal{R}^{\mathrm d}$ and model packet routing paths $\mathcal{R}^{\mathrm m}$;
    \STATE Calculate $T_i^{\mathrm{total}}$ according to~\eqref{T_total};
    \STATE Update $\alpha_i^{\mathrm{low}}$ and $\alpha_i^{\mathrm{up}}$ according to the BS criterion;
\ENDFOR
\STATE \textbf{return} $\alpha_i$, $\mathcal{R}^{\mathrm d}$, $\mathcal{R}^{\mathrm m}$, and $T_i^{\mathrm{total}}$.
\end{algorithmic}
}
\end{algorithm}

\section{Proposed Solution}
\label{sec:opt_alg} 

\subsection{Offline Policy Profiling}
To efficiently solve the subproblem~\eqref{sub_problem_2}, which is transformed as a Dec-POMDP, the MARL routing policy is trained offline. 
Let $\mathcal{N}_{\mathrm a}\subseteq\mathcal{N}$ denote the set of active routing agents.
All agents share the same utility network architecture with parameter vector $\phi$, and the individual action-value function of agent $n\in\mathcal{N}_{\mathrm a}$ at time $t$ is denoted by $Q_n(\tau_n^t,a_n^t;\phi)$, where $\tau_n^t$ is the local action-observation history of agent $n$ up to time $t$.
To enhance cooperation among agents, a mixing network is introduced to aggregate the individual action values into a joint action-value function
\begin{equation}
    Q_{\mathrm{total}}(\boldsymbol{\tau}_t,\mathbf{a}_t,\mathbf{x}_t;\phi,\psi)=f_{\mathrm{mix}}(Q_1,\ldots,Q_{N_{\mathrm a}},\mathbf{x}_t;\psi),
\end{equation}
where $\mathbf{x}_t$ denotes the global state and $\psi$ is the parameter vector of the mixing network. Following the QMIX framework, the optimal joint action can be obtained from the individual greedy actions under the monotonicity constraint
\begin{equation}
    \frac{\partial Q_{\mathrm{total}}(\boldsymbol{\tau}_t,\mathbf{a}_t,\mathbf{x}_t;\phi,\psi)}{\partial Q_n(\tau_n^t,a_n^t;\phi)}\ge 0,\quad \forall n\in\mathcal{N}_{\mathrm a}.
\end{equation}
The MARL routing model is trained by minimizing the temporal difference loss, which can be expressed as
\begin{equation}
\label{eq:loss}
    L(\phi,\psi)=\mathbb{E}\!\left[\left(y_t-Q_{\mathrm{total}}(\boldsymbol{\tau}_t,\mathbf{a}_t,\mathbf{x}_t;\phi,\psi)\right)^2\right].
\end{equation}
The target value is defined as
\begin{equation}
    y_t=r_t+\eta\,Q_{\mathrm{total}}(\boldsymbol{\tau}_{t+1},\mathbf{a}_{t+1}^{\star},\mathbf{x}_{t+1};\phi^{-},\psi^{-}),
\end{equation}
where $\eta$ is the discount factor, $(\phi^{-},\psi^{-})$ are the parameters of the target networks, and $\mathbf{a}_{t+1}^{\star}$ denotes the greedy joint action at time $t+1$, i.e.,
\begin{equation}
    \label{argmaxa}
    \mathbf{a}_{t+1}^{\star}
    =
    \left(
    \arg\max_{a_n}Q_n(\tau_n^{t+1},a_n;\phi)
    \right)_{n\in\mathcal{N}_{\mathrm a}}.
\end{equation}

\subsection{Online Optimization}
{
The purpose of online optimization is to determine the task allocation ratio via the BS algorithm and obtain the routing strategy using the trained MARL policy. 
The basic principle is to balance the delay between the remote sensing satellite and the computing satellite.
Specifically, for each allocation ratio searched in a BS iteration, the service delay is evaluated by solving the offloading subproblem and the routing subproblem. 
If the small model processing delay is larger than the sum of the large model processing delay and the transmission delay, the BS algorithm increases the allocation ratio; otherwise, it decreases.
Through this online procedure, the allocation ratio and the routing strategy are optimized in an alternating manner.
The online optimization terminates when the service delay obtained in the current iteration is larger than that obtained in the previous iteration.
After multiple iterations, the delays of the remote sensing satellite and the computing satellite gradually become balanced, which enables efficient utilization of the computational resources.
The overall BS-MARL solution procedure is summarized in \hyperref[alg1]{\textbf{Alg. 1}}.
}

\section{Performance Evaluation}\label{sec:sim_res}
\subsection{Simulation Setup}
We consider a Walker constellation constructed in Satellite Tool Kit, consisting of $8$ orbital planes with $8$ satellites in each plane, deployed at an altitude of $800~\text{km}$ and an inclination of $30^\circ$. In the simulations, a remote sensing image classification task is adopted as the representative application scenario. The simulation runs from $00{:}00{:}00$ to $12{:}00{:}00$ on October~18, 2025, with a slot duration of $30$ seconds, during which the constellation is treated as quasi-static. The system includes $6$ remote sensing satellites and $3$ computing satellites.

To support collaborative processing, the remote sensing satellite deploys a ResNet18 model with $11.7~\text{M}$ parameters, while the computing satellite deploys a ResNet101 model with $45~\text{M}$ parameters~\cite{ZSH-2025}. 
The transmission power is set to $10~\mathrm{W}$, and the transmit and receive antenna gains are both set to $30~\mathrm{dBi}$. The carrier frequency is $23~\mathrm{GHz}$, and the link margin is $1.5~\mathrm{dB}$. The residual data quantization is set to $0.55$~bit/pixel.
The computing capacities of the remote sensing satellite and the computing satellite are set to $1$~TOPS and $10$~TOPS, respectively, with a hardware utilization factor of $0.8$.
The following benchmarks are considered for comparison:

\begin{itemize}
    \item  Small model processing scheme: all tasks are processed locally on the remote sensing satellite. 
    \item Centralized large model processing scheme: all frames are packaged into data packets and transmitted via ISLs to the computing satellite for target classification.
    \item Even split processing scheme: the proposed large-small model collaboration scheme is employed, $\alpha_{i} = 0.5$.
\end{itemize}

\begin{figure*}[t]
    \centering

    \begin{minipage}[b]{0.24\textwidth}
        \centering
        \includegraphics[width=\textwidth]{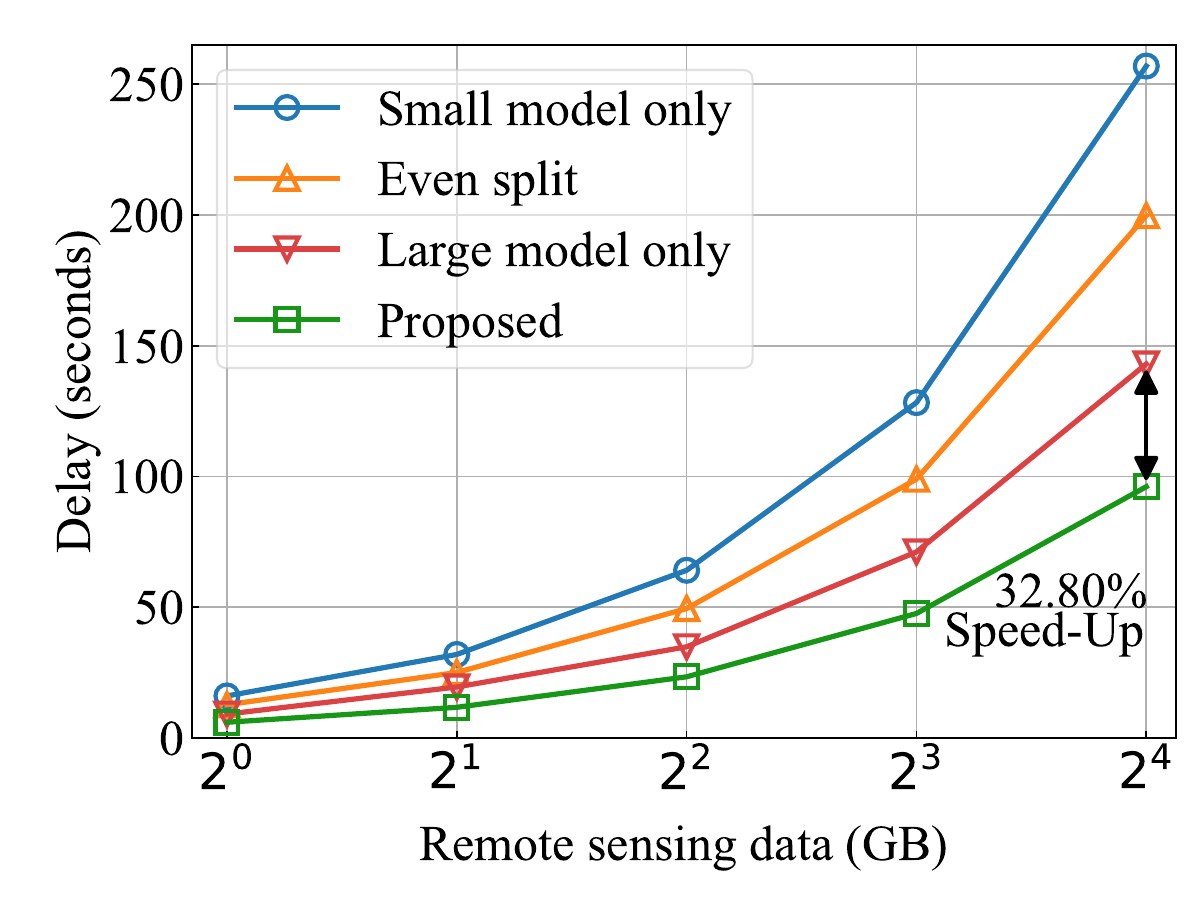}
        \captionof{figure}{Service delay under varying remote sensing data.}
        \label{delay_volume}
    \end{minipage}
    \hfill
    \begin{minipage}[b]{0.24\textwidth}
        \centering
        \includegraphics[width=\textwidth]{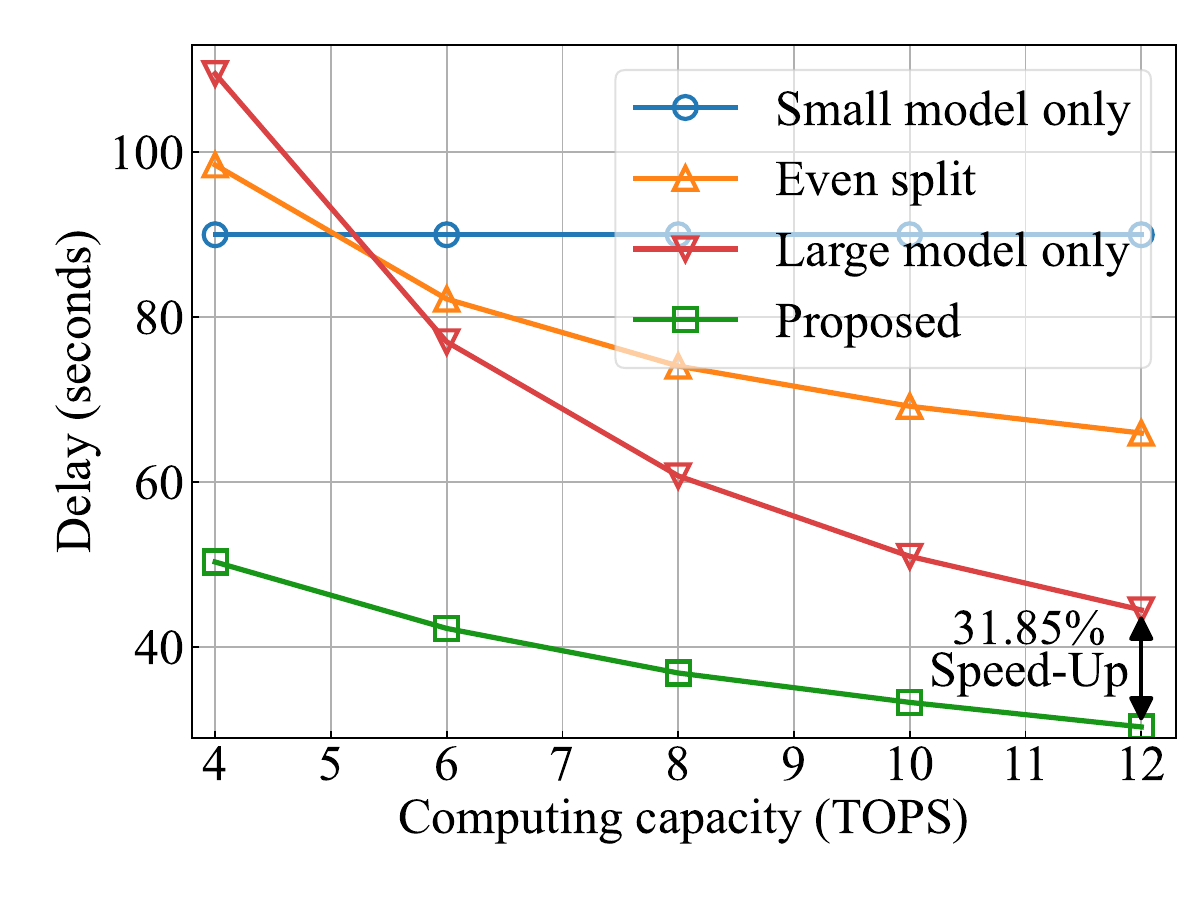}
        \captionof{figure}{Service delay with respect to computing capacity.}
        \label{delay_computing_capable}
    \end{minipage}
    \hfill
    \begin{minipage}[b]{0.24\textwidth}
        \centering
        \includegraphics[width=\textwidth]{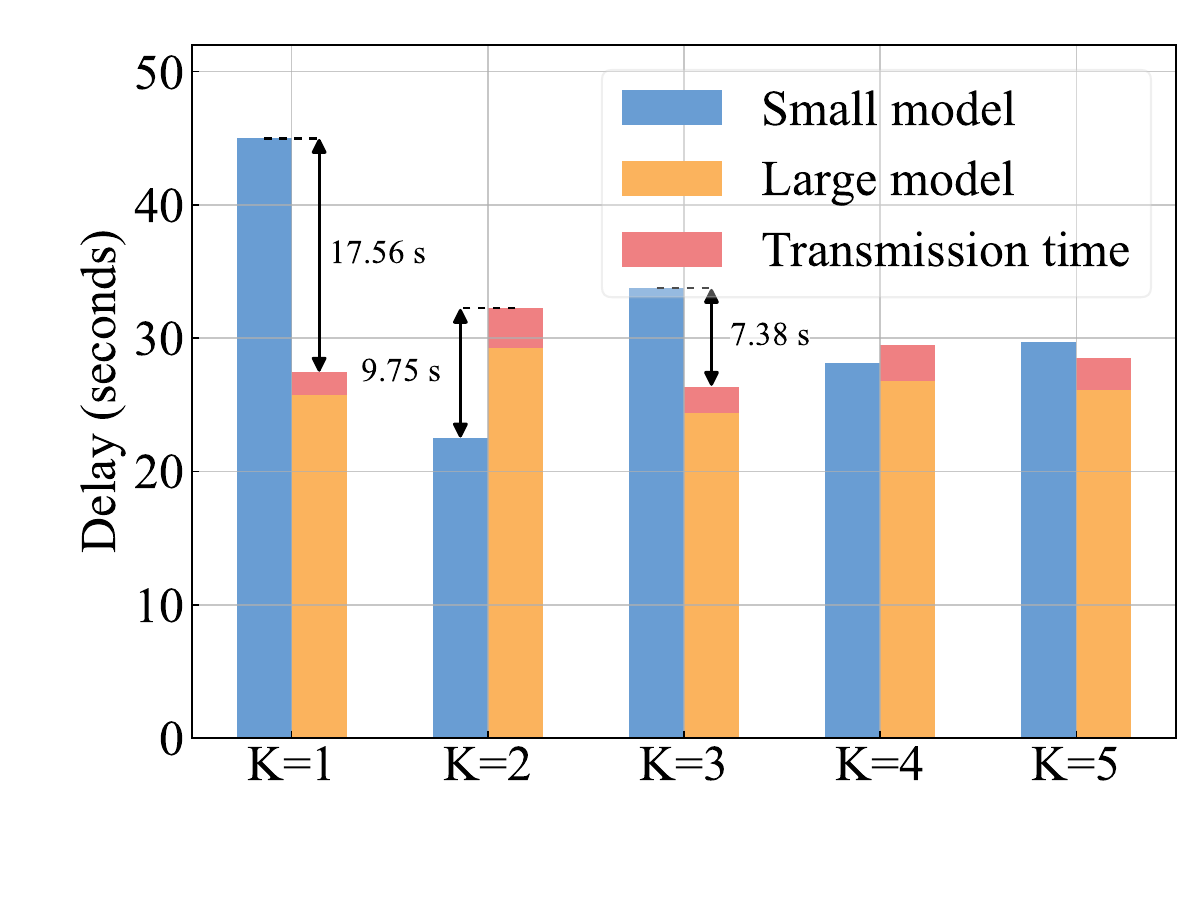}
        \captionof{figure}{Service delay under different bisection iterations.}
        \label{delay_bar}
    \end{minipage}
    \hfill
    \begin{minipage}[b]{0.24\textwidth}
        \centering
        \includegraphics[width=\textwidth]{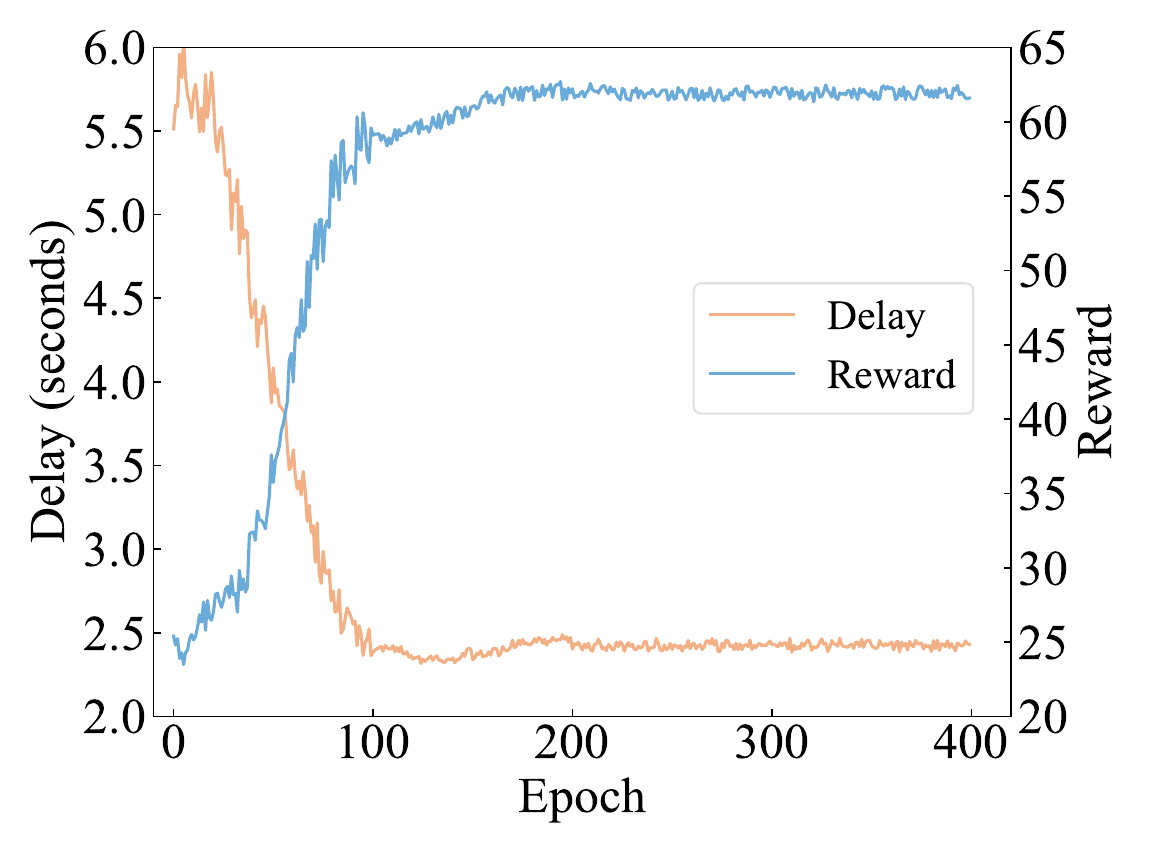}
        \captionof{figure}{Transmission delay and reward over epochs.}
        \label{tra_delay}
    \end{minipage}
\end{figure*}

\subsection{Simulation Results}
As shown in Fig.~\ref{delay_volume}, the service delay of all four schemes with different data volumes is illustrated. 
The proposed scheme adaptively adjusts the allocation ratio $\alpha_{i}$ based on the satellite network's load conditions and the multiplexing of ISLs, thereby reducing the average service delay by about 32\% relative to other schemes.
The small model processing framework performs the worst due to the limited onboard computing resources of the remote sensing satellite.

Figure~\ref{delay_computing_capable} shows the delays of four different schemes under varying computing capacities on computing satellites.
The proposed large-small model collaboration scheme delivers the best performance, reducing the service delay by at least 31\%.
As the computing capacity of the computing satellite increases, the allocation ratio correspondingly rises, enabling more frames to be processed by the large model. 
As a result, the even split processing scheme performs better at low computing capacities, while the centralized large model processing scheme becomes superior at high computing capacities.

In Fig.~\ref{delay_bar}, we illustrate how the proposed algorithm iteratively adjusts the allocation ratio via the BS algorithm to reduce the overall delay.
With $\alpha_{i}$ initialized to $0.5$, the computing satellite experiences an idle time of $17.56~\text{s}$. 
The BS algorithm therefore increases $\alpha_{i}$ in the next iteration.
As the iterations proceed, the relative processing delays of the remote sensing satellite and the computing satellite may alternate, and the gap between them gradually becomes negligible, thereby leading to more efficient utilization of computing resources.

Figure~\ref{tra_delay} shows the offline MARL routing policy training process. As the number of training epochs increases, the agents gradually learn effective routing policies, and the oscillation amplitude of the curves progressively decreases, leading to the convergence of both transmission delay and reinforcement learning reward. Moreover, the proposed BS-MARL algorithm consistently converges within 100 epochs under different packet sizes and network environments, demonstrating the strong robustness of the overall framework.

\section{Conclusion}
\label{sec:conclusion}
In this paper, we have proposed a large-small model collaboration scheme for LEO satellite networks, which can efficiently reduce service delay via offloading decisions and routing strategies.
Additionally, we have formulated a service delay minimization problem and then proposed a BS-MARL algorithm to solve it.
The proposed scheme can facilitate the practical deployment of onboard computing in LEO satellite networks with heterogeneous computational resources.
For future work, we will focus on knowledge distillation-based model updating in the large-small model collaboration scheme.



\bibliographystyle{IEEEtran}
\bibliography{referencebox}

\end{document}